\documentclass[letterpaper]{article} 
\usepackage{aaai25}  
\usepackage{times}  
\usepackage{helvet}  
\usepackage{courier}  
\usepackage[hyphens]{url}  
\usepackage{graphicx} 
\usepackage{natbib}  
\usepackage{caption} 
\usepackage{algorithm}
\usepackage{algorithmicx}

\usepackage{algpseudocode}

\usepackage{microtype}
\usepackage{epsfig}

\usepackage{caption}
\usepackage{float}
\usepackage{placeins}
\usepackage{color, colortbl}
\usepackage{enumitem}
\usepackage{tabularx}
\usepackage{xstring}
\usepackage{multirow}
\usepackage{xspace}
\usepackage{url}
\usepackage{subcaption}
\usepackage{xcolor}
\usepackage[hang,flushmargin]{footmisc}

% ...

\newcommand{\R}[1]{{%
    \textbf{%
        \ifstrequal{#1}{1}{\textcolor{red}{R#1}}{%
        \ifstrequal{#1}{2}{\textcolor{blue}{R#1}}{%
        \ifstrequal{#1}{3}{\textcolor{magenta}{R#1}}{%
        \ifstrequal{#1}{4}{\textcolor{teal}{R#1}}{%
                           \textcolor{cyan}{R#1}%
        }}}}%
    }%
}}

\newcommand{\tempcolor}{black}

\newcommand{\cutabstractup}{\vspace*{-0.00in}}
\newcommand{\cutabstractdown}{\vspace*{-0.00in}}

\newcommand{\cutsectionup}{\vspace*{-0.00in}}
\newcommand{\cutsectiondown}{\vspace*{-0.00in}}

\newcommand{\cutsubsectionup}{\vspace*{-0.00in}}
\newcommand{\cutsubsectiondown}{\vspace*{-0.01in}}

\usepackage{newfloat}
\usepackage{listings}
\DeclareCaptionStyle{ruled}{labelfont=normalfont,labelsep=colon,strut=off} 
\floatstyle{ruled}
\newfloat{listing}{tb}{lst}{}
\floatname{listing}{Listing}
\title{Step-Calibrated Diffusion for Biomedical Optical Image Restoration}
\author{
    Yiwei Lyu\textsuperscript{\rm 1}\equalcontrib,Sung Jik Cha\textsuperscript{\rm 2}\equalcontrib,Cheng Jiang\textsuperscript{\rm 1},Asadur Zaman Chowdury\textsuperscript{\rm 1},Xinhai Hou\textsuperscript{\rm 1}\\
    Edward S. Harake\textsuperscript{\rm 1},Akhil Kondepudi\textsuperscript{\rm 1},Christian Freudiger\textsuperscript{\rm 3},Honglak Lee\textsuperscript{\rm 1}\textsuperscript{\rm 4},Todd C. Hollon\textsuperscript{\rm 1}
}
\affiliations{
    \textsuperscript{\rm 1}University of Michigan,
    \textsuperscript{\rm 2}Western Michigan University,
    \textsuperscript{\rm 3} Invenio Imaging,
    \textsuperscript{\rm 4} LG AI Research\\

    yiweilyu@umich.edu
}

\begin{document}

\maketitle

\cutabstractup
\begin{abstract}

High-quality, high-resolution medical imaging is essential for clinical care. Raman-based biomedical optical imaging uses non-ionizing infrared radiation to evaluate human tissues in real time and is used for early cancer detection, brain tumor diagnosis, and intraoperative tissue analysis. Unfortunately, optical imaging is vulnerable to image degradation due to laser scattering and absorption, which can result in diagnostic errors and misguided treatment. Restoration of optical images is a challenging computer vision task because the sources of image degradation are multi-factorial, stochastic, and tissue-dependent, preventing a straightforward method to obtain paired low-quality/high-quality data. Here, we present Restorative Step-Calibrated Diffusion (RSCD), an unpaired diffusion-based image restoration method that uses a step calibrator model to dynamically determine the number of steps required to complete the reverse diffusion process for image restoration. RSCD outperforms other widely used unpaired image restoration methods on both image quality and perceptual evaluation metrics for restoring optical images. Medical imaging experts consistently prefer images restored using RSCD in blinded comparison experiments and report minimal to no hallucinations. Finally, we show that RSCD improves performance on downstream clinical imaging tasks, including automated brain tumor diagnosis and deep tissue imaging. Our code is available at \url{https://github.com/MLNeurosurg/restorative_step-calibrated_diffusion}. 

\cutabstractdown
\cutabstractdown
\end{abstract}

\begin{figure}[t!]
    \centering
    \vspace{-2pt}
    \includegraphics[width=0.92\columnwidth]{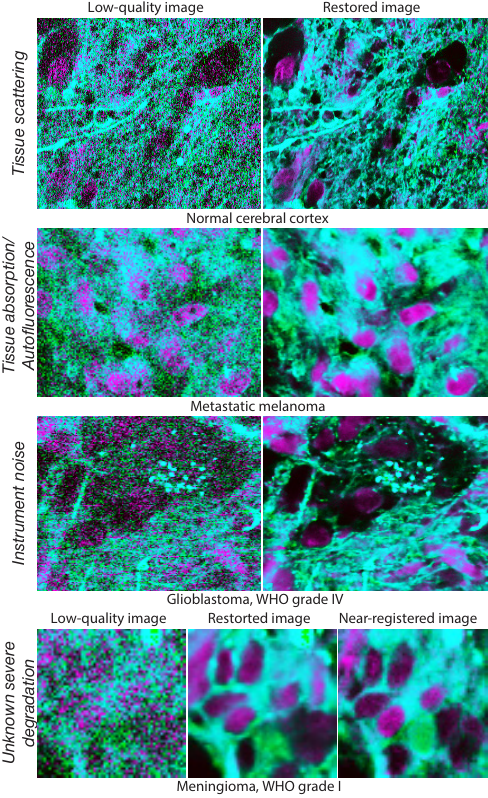}
    \vspace{-3pt}
    
    \caption{Examples of biomedical optical images restored using our proposed method, \textbf{\emph{RSCD}}. Across a range of known and unknown sources of image degradation, RSCD provides high-quality image restoration of fresh, surgical specimens imaged during brain tumor surgery. Our method can restore optical images with severe image degradation such that, after restoration, they can be used for downstream clinical tasks, including automated brain tumor diagnosis and deep tissue imaging during surgery.}
    
    \label{fig:fig1}
\end{figure}
\vspace{-10pt}

\cutsectionup
\section{Introduction}
\label{sec:intro}
\cutsectiondown

\begin{figure*}
    \centering
    \includegraphics[width=\textwidth]{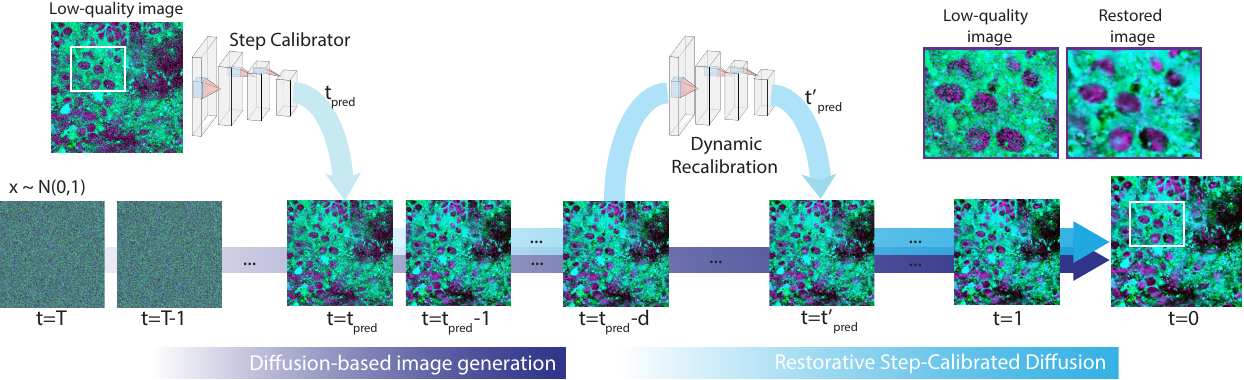}
    \vspace{-5mm}
    \caption{An overview of \emph{\textbf{Restorative Step-Calibrated Diffusion (RSCD)}}. We view the low-quality image as the output of an incomplete diffusion generation process that starts from Gaussian noise ($t=T$) and performs $T$ steps of denoising (reverse diffusion) to generate a restored image at $t=0$. We use a step calibrator model to predict $t_{pred}$, the number of steps of diffusion model denoising needed for image restoration, and we perform the reverse diffusion starting from $t_{pred}$. In addition, we use dynamic recalibration to dynamically adjust the number of steps required for optimal image restoration, $t'_{pred}$.    
The dynamic recalibration process and subsequent $d$ steps of reverse diffusion denoising can be repeated until the restoration process runs to completion, obtaining the restored image at $t=0$.
}
    \label{fig:methodfig}
\vspace{-0.15in}
\end{figure*}

Medical imaging plays a major role in clinical medicine. Computed tomography, radiography, magnetic resonance imaging, and optical imaging are examples of common and indispensable medical imaging modalities used for diagnosis, guiding treatment decisions, and monitoring treatment response. Raman-based biomedical optical imaging uses Raman scattering to non-invasively evaluate human tissues for diagnostic purposes. As an advanced medical imaging modality, it has an increasing role in patient care and is now being used for non-invasive cancer detection \cite{Waterhouse2019-wj, Lui2012-xm}, brain tumor diagnosis \cite{Hollon2020-ez, Hollon2023-uf}, and surgical specimen analysis \cite{Orringer2017-nn, Mannas2023-qd, Hoesli2017-yp}. Raman-based optical imaging has several advantages over conventional medical imaging as it does not require ionizing radiation, does not cause tissue damage, and can acquire images rapidly (within seconds) at the patient's bedside. Optical imaging uses light in the infrared electromagnetic spectrum to visualize biological tissues. Unfortunately, imaging within this spectral region causes optical imaging to be vulnerable to image degradation. Moreover, the sources of image degradation are multifaceted \cite{Waterhouse2019-wj, Manifold2019-ft} (see Figure \ref{fig:fig1}):

\vspace{-0.05in}
\begin{itemize}[noitemsep]
    \item \emph{Tissue scattering}: scattering of the incident light can result in noisy images, especially when imaging at depth.
    \item \emph{Tissue absorption}: biologically tissues contain chromophores that absorb light at specific wavelengths, reducing signal strength.
    \item \emph{Auto-fluorescence}: some biological tissues have intrinsic fluorescence that degrades optical images by reducing signal-to-noise ratios. 
    \item \emph{Instrument noise}: image noise can be introduced from the detectors, electronic components, and external sources of interference.
\end{itemize}

\vspace{-0.05in}
\FloatBarrier



The frequency and degree of image degradation are inherently unpredictable, making widespread clinical integration of Raman-based optical imaging challenging \cite{Hollon2016-uk, Waterhouse2019-wj}. Moreover, while small-scale imperfectly-paired image data has been generated in controlled laboratory settings \cite{Manifold2019-ft, Weigert2018-jr}, it is impossible to collect a large-scale paired dataset of perfectly aligned low-quality/high-quality clinical optical imaging that includes the full range of possible image degradations~\cite{min2011coherent,audier2020noise,moester2015optimized}.

Unfortunately, previous unpaired restoration methods suffer from hallucinations and perceptually poor reconstructions \cite{Belthangady2019-ax}, which can have detrimental downstream effects in medical imaging, increasing the risk of nondiagnostic images or diagnostic errors. An ideal method for Raman-based optical image restoration would (1) restore images degraded from a wide range of corruption sources, (2) only require unpaired data, (3) avoid hallucinations or perceptual artifacts, and (4) be time-efficient to allow for real-time, intraoperative image restoration.

We present \emph{\textbf{Restorative Step-Calibrated Diffusion (RSCD)}}, a novel diffusion-based unpaired image restoration method that efficiently restores low-quality Raman-based optical images with minimal to no perceptual artifacts or hallucinations. \textcolor{\tempcolor}{The rationale behind RSCD is that (1) Gaussian-based DDPMs have demonstrated generalization capacities in restoring non-Gaussian degradation~\cite{chung2022come,Chung2023-qt}, which makes them the ideal restoration model for unpredictable and unknown degradations; (2) DDPM restoration by directly performing reverse diffusion steps on degraded images requires much fewer steps than the full generative reverse diffusion process; and finally (3) when the degradation is unpredictable, the number of reverse diffusion steps required should vary for each image depending on the severity and pattern of the degradation.}

 \textcolor{\tempcolor}{Thus, RSCD includes a step calibrator model that determines the number of restoration steps required and a generative diffusion model that completes the restorative steps.}
Both models can be trained using unpaired high-quality images. RSCD is hallucination-resistant because it only involves editing the image via noise removal rather than full image generation from a random prior as is conventionally done in generative diffusion-based image restoration methods \cite{Kawar2022-vc}. Moreover, RSCD is time-efficient because it does not require the full reverse diffusion process. To further improve restoration quality and stability on unpredictable degradations, we designed a dynamic recalibration process that dynamically adjusts the number of remaining steps \emph{during} restoration. 
The well-trained step calibrator model and dynamic recalibration enable RSCD to consistently restore biomedical optical images that have various degrees and distributions of image degradation, as shown in Figure \ref{fig:fig1}.

We summarize our contributions as follows:
\vspace{-0.03in}
\begin{itemize}[noitemsep]
    \item We propose a novel unpaired image restoration method, RSCD, that is fast, reliable, and ideally suited for Raman-based optical imaging where noise is unpredictable with varying strength and pattern across and within images.
    \item RSCD is evaluated against other image restoration baselines, and outperforms them on image quality metrics. RSCD also achieves state-of-the-art performance on various unpaired perceptual metrics.
    \item Optical imaging experts consistently prefer images restored via RSCD over other methods, and report minimal to no hallucinations during human evaluations.
    \item RSCD can improve performance on downstream clinical computer vision tasks, including automated brain tumor diagnosis and deep tissue imaging.
\end{itemize}

\cutsectionup
\section{Background}
\label{sec:background}

\cutsubsectionup
\subsection{Intraoperative Raman-based optical imaging}
\cutsubsectiondown
In this paper, we focus on the restoration of stimulated Raman histology (SRH), a rapid and label-free optical imaging method based on Raman spectroscopy \cite{Freudiger2008-gj}. SRH is used for a wide range of biomedical imaging tasks and has been clinically validated for imaging fresh, unprocessed human tissues and surgical specimens \cite{Orringer2017-nn}. SRH begins with imaging a tissue specimen at two Raman shifts, 2845cm\textsuperscript{-1} and 2930cm\textsuperscript{-1}, which highlights the optical image features generated by the lipid and protein concentrations, respectively, to generate image contrast. SRH can capture high-resolution, diagnostic-quality images across multiple organs and tissues \cite{Orringer2017-nn, Hollon2018-hl}. A major advantage of SRH is that it can be performed rapidly ($\sim$1 minute) without the need for tissue processing or staining, making it ideally suited for intraoperative tissue evaluation and diagnosis during surgery.

\cutsubsectionup
\subsection{Generative diffusion models}
\cutsubsectiondown

\label{sec:diffusionbackground}
Denoising diffusion probabilistic models are used to generate high-quality and diverse images \cite{Ho2020-tm, Dhariwal2021-mh}. Training diffusion models consist of two processes. The first is the forward diffusion process, which gradually adds Gaussian noise to an image $x_0$ over $T$ steps to obtain $x_1,x_2,\dots,x_T$, where $x_t \sim N(\sqrt{1-\beta_t} x_{t-1}; \beta_tI)$
for each $1\leq t \leq T$ and $\beta_1, \dots, \beta_T$ follows a noise schedule. The total noise added over the $T$ steps should be strong enough to reduce the image to Gaussian noise, such that $x_T \sim N(0,I)$. Since combining multiple steps of Gaussian noise results in Gaussian noise, we see that for each $t$, $x_t \sim N(\sqrt{\bar{\alpha}_t}x_0,(1-\bar{\alpha}_t)I)$ where $\alpha_t=1-\beta_t$ and $\bar{\alpha}_t=\Pi_{i=1}^t\alpha_i$. The second process is the restoration phase, where we train a model $\epsilon_\theta$ that takes in a noised image ($x_t=\sqrt{\bar{\alpha}_t}x_0+\sqrt{1-\bar{\alpha}_t}\epsilon$ where $\epsilon \sim N(0,I)$) and the step number $t$, and tries to predict the noise ($\epsilon$), and the training loss is $||\epsilon - \epsilon_\theta(x_t,t)||$. When we use the model to generate an image, we start by randomly sampling $x_T\sim N(0,1)$, and we gradually remove noise by sampling $x_{t-1} \sim N\left(\frac{1}{\sqrt{\alpha_t}}(x_t-\frac{1-\alpha_t}{\sqrt{1-\bar{\alpha_t}}}\epsilon_\theta(x_t,t)),\beta_tI\right)$, and we repeat this $T$ times until we reach clean image $x_0$.


\cutsectionup
\section{Methodology}
\label{sec:method}

\begin{figure}[t]
    \centering
    \includegraphics[width=1.0\columnwidth]{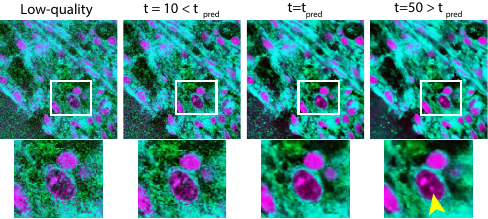}
    \vspace{-0.2in}
    \caption{Importance of the \textbf{Step calibrator}. If we perform less than the optimal number of diffusion steps ($t < t_{pred}$), the image remains degraded. If we perform more than the needed steps ($t > t_{pred}$), the output image is excessively smooth, fine details are removed, and contains hallucinations (yellow arrow). ($t_{pred}=34$ for this image.)}
    \label{fig:stepillus}
\vspace{-0.1in}
\end{figure}



\textcolor{\tempcolor}{The key idea behind RSCD is that when using the last $t$ steps of a $T$-step generative diffusion process to perform image restoration, the number of steps $t$ required should differ for each image due to variances in degradation severity and pattern. RSCD uses a trainable step calibrator model that predicts the value of $t$ for each low-quality image. The step calibration is crucial for restoration quality because, as illustrated in Figure~\ref{fig:stepillus}, if $t$ is too small, then the reverse diffusion process cannot complete a sufficient number of steps to restore the low-quality image. If $t$ is too large, then the reverse diffusion process generates excessively smooth and homogenized images with hallucinations. Thus, setting a fixed $t$ value as a hyperparameter results in suboptimal image quality and perceptual features.}
 To further
improve restoration quality, we dynamically recalibrate the
number of steps needed to restore the image during the reverse diffusion process. An overview of our method is shown in Figure~\ref{fig:methodfig}.

\cutsubsectionup
\subsection{Training Data}
\label{sec:data}
\cutsubsectiondown

Training data was generated from approximately 2500 patients who underwent intraoperative SRH imaging to evaluate tissue during surgery \cite{Orringer2017-nn}. Whole slide SRH images are approximately 6000$\times$6000 pixels, which are then divided into 256$\times$256 pixel patches, resulting in $\sim$1 million total patches. To obtain high-quality SRH images, optical imaging experts manually selected 4.5K high-quality patches, and then we automatically filtered through the remaining patches to obtain 840K relatively high-quality patches, using the 4.5K patches as guidance. Details of the filtering process are included in Appendix~\ref{app:auto}.

\cutsubsectionup
\subsection{Step calibrator}
\cutsubsectiondown



We used a ResNet-50 model with MLP prediction head as the step calibrator. During training, a step number $t \sim \mathcal{U}(0,T)$ is sampled and $t$ steps of Gaussian noise are added to a high-quality image according to a cosine schedule \cite{Ho2020-tm}. The calibrator is trained to predict $t$ using an $L_2$ loss between the prediction $t_{pred}$ and $t$. A challenge we identified with this naive implementation is that the severity of image degradation varies \emph{within} an optical image, and the trained model tends to predict steps based on the region with the least amount of degradation. To address this issue, we perform the following augmentation when training the step calibrator: we randomly divide the image into two regions, and we add $t$ steps of noise to one of the regions; we then sample a second, smaller $t'$ such that $0\leq t' \leq t$, and add $t'$ steps of noise to the other region. We train the model to predict the larger noise $t$ to ensure the step calibrator favors calibrations that will restore the most degraded regions in the image. We show an example of this augmentation in Figure~\ref{fig:aug} and the full algorithm in Algorithm~\ref{alg:sp}, both in Appendix~\ref{app:steppred}.

In practice, we set $T=1000$. After training, the step calibrator was used to predict noise level on approximately 63K low-quality images that had the lowest score from section~\ref{sec:data}. The distribution of the predicted $t_{pred}$ is shown in Figure~\ref{fig:histogram} in Appendix~\ref{app:steppred}. We found that the majority of low-quality images require less than 100 steps to restore, and the number of low-quality images decreases exponentially as the number of steps increases. No low-quality image required more than 200 steps for image restoration. 


\cutsubsectionup
\subsection{Diffusion model}
\cutsubsectiondown

The diffusion model is trained as a generative diffusion model that generates high-quality SRH images unconditionally with a cosine noise schedule. We follow a similar training objective and procedure as described in \cite{Ho2020-tm}, except that we use a shortcut for training efficiency: RSCD only requires training the model with noise level randomly sampled between 1 and $T'$ steps (where $T'$ just need to be larger than the maximum $t_{pred}$ of low-quality images) instead of sampling from the full range 1 to $T$ when training diffusion-based generation models, as shown in Algorithm~\ref{alg:train} in Appendix~\ref{app:train}. In practice, we set $T'=200$ due to the distribution in of $t_{pred}$ as discussed in the previous section. This allows us to train the model more efficiently than conventional diffusion-based image restoration methods \cite{Kawar2022-vc}. We first train the model for one pass through all 840K images, then fine-tune the model on the 4.5K high-quality images for 20 epochs.

\cutsubsectionup
\subsection{Dynamic Recalibration}
\cutsubsectiondown

The step calibrator is trained to predict Gaussian noise levels; however, Raman-based optical image degradation is not limited to Gaussian noise. When image degradation deviates significantly from Gaussian noise, the step calibrator is more likely to under or overestimate the required number of diffusion steps, which results in poor image restorations. Therefore, to better calibrate the number of steps \emph{during} the image restoration process, we perform dynamic recalibration: after we predict $t_{pred}$ for an input low-quality image, instead of directly performing all $t_{pred}$ steps of denoising, we only apply $d$ steps of denoising, i.e. $x_{t_{pred}} \rightarrow x_{t_{pred}-1} \rightarrow ... \rightarrow x_{t_{pred}-d}$. Then, we will use the step calibrator to predict the remaining steps of denoising needed for $x_{t_{pred}-d}$, and continue denoising starting from the updated predicted number of steps. Another $d$ steps of denoising are performed before additional calibration. We repeat the process until the predicted number of steps remaining is less than $d$, and then we complete the remaining steps without additional recalibration. The process is illustrated in Algorithm~\ref{alg:dr}. 

\textcolor{\tempcolor}{Note that even though the observed noise from SRH images is not Gaussian, we can train our step calibrator and diffusion model with Gaussian noise because (1) Gaussian diffusion models are known to have generalization capacities for restoring degradations that deviates from Gaussian distributions~\cite{Chung2023-qt,chung2022scorebaseddiffusionmodelsaccelerated}, as small Gaussian noise is added during each DDPM step that makes the resulting degradation distribution more Gaussian-like; and (2) dynamic recalibration can mitigate possible step prediction inaccuracies due to different noise patterns by dynamically adjusting the remaining number of steps.}


\begin{algorithm}[t]
\caption{\small Restorative Step-Calibrated Diffusion with \textcolor{red}{Step Calibration} and \textcolor{blue}{Dynamic Recalibration}: Sampling}\label{alg:dr}
\begin{algorithmic}[1]
\small
\State \textbf{Requires:} Low-quality image $x$, \textcolor{red}{step calibrator $S$}, diffusion model $\epsilon_\theta$, \textcolor{blue}{recalibration interval $d$}, total steps $T$, hyperparameters based on noise schedule $\alpha_1$,...,$\alpha_T$,$\bar{\alpha}_1$,...,$\bar{\alpha}_T$,$\beta_1$,...,$\beta_T$
\color{red}
\State $t \gets S(x)$ \textcolor{gray}{\emph{\# Initial step calibration}}
\color{black}
\State $t_{end} \gets 0 $ 
\While{$t > 0$} 

\State $x_t \gets x$ 
\color{blue}
\State $t_{end} = max(t-d,0)$ \textcolor{gray}{\emph{\# d-steps of reverse diffusion}} 
\color{black}
\For{$k$ \textbf{in} $[t, t-1, ..., t_{end}+1]$}
\State $z\sim N(0,1)$
    \State $x_{k-1}\gets \frac{1}{\sqrt{\alpha_k}}(x_k-\frac{1-\alpha_k}{\sqrt{1-\bar{\alpha_k}}}\epsilon_\theta(x_k,k))+\sqrt{\beta_k}z$
\EndFor
\State $x \gets x_{t_{end}}$
\color{blue}
\State $t \gets S(x)$ \textcolor{gray}{\emph{\# Dynamic recalibration}}
\color{black}
\EndWhile
\State \Return $x$
\end{algorithmic}
\end{algorithm}
\cutsectionup
\section{Experiments}

\cutsubsectionup
\subsection{Baselines and Ablations}
\cutsubsectiondown

In the following experiments, we compare RSCD to the several unpaired image restoration baselines: CycleGAN\cite{Zhu2017-wf}, synthetic noise/noise2noise \cite{Lehtinen2018-zd, Manifold2019-ft}, conditional diffusion~\cite{Kawar2022-vc, Saharia2022-co}, CCDF~\cite{chung2022come}, regularized reverse diffusion (RRD) \cite{Chung2023-qt}, deep image prior~\cite{Ulyanov2016-kz}, and median blur. In addition, we also conduct ablation studies over the following design choices: dynamic recalibration, step calibrator, and cosine noise schedule. We compare our method against no dynamic recalibration, no step calibrator (either using a non-parametric noise estimator \cite{Chen2015-ij} as replacement or denoise for a fixed number of steps), evaluating the commonly-used linear noise schedule versus the cosine schedule, and training our step calibrator without augmentation. Details of each baseline and ablation can be found in Appendix~\ref{app:baseline}.

\begin{figure*}[t]
    \centering
    \includegraphics[width=\textwidth]{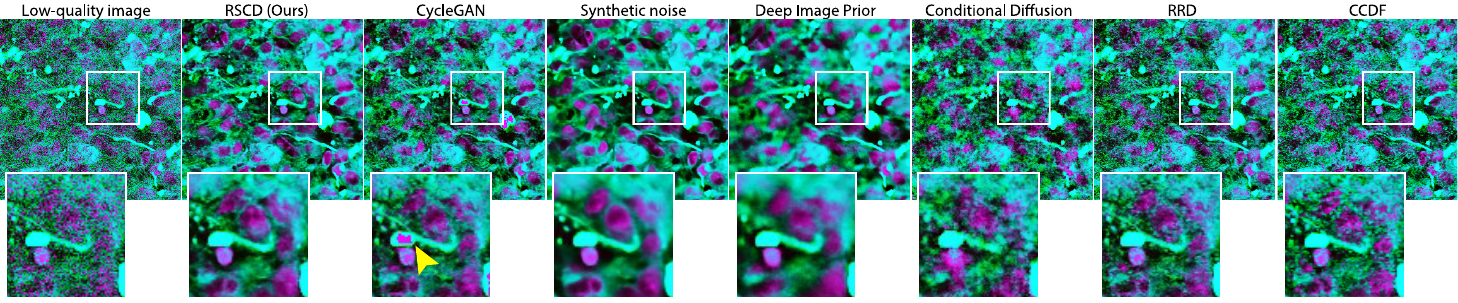}
    \vspace{-7mm}
    \caption{Visual comparison of unpaired image restoration methods. CycleGAN hallucinates/inpaints nuclei within non-cellular structures (yellow arrow). Synthetic noise and Deep Image Prior (DIP) produce overly smoothed, unrealistic images. Conditional diffusion and Regularized Reverse Diffusion (RRD) generally perform insufficient image restoration.}
    \label{fig:compare}
    \vspace{-0.1in}
\end{figure*}

\cutsubsectionup
\subsection{Evaluation on Unpaired Images}
\label{sec:unpaired}
\cutsubsectiondown

We evaluate the perceptual quality of image restoration on 12K unpaired low-quality images sampled from the largest public SRH dataset, OpenSRH \cite{Jiang2022-cj}. We evaluate RSCD against the above baseline methods and ablations. We use Frechet Inception Distance (FID) \cite{Heusel2017-bw} and CLIP Maximum Mean Discrepancy (CMMD) \cite{jayasumana2024rethinking} between the restored images and the 4.5K expert-selected high-quality images as metrics.

We report all results in Table~\ref{tab:unpaired}. RSCD outperformed all baselines and ablations, indicating that our method produces high-quality and realistic image restorations of low-quality Raman-based optical imaging. We present an example of qualitative comparison in Figure~\ref{fig:compare}, with additional examples in Figure~\ref{fig:unpairedadditional} in Appendix. We provide additional details on training/inference efficiency and compute resources in Appendix~\ref{app:compres}.

\begin{table}[t]
\centering
\scalebox{0.84}[0.84]{
\begin{tabular}{ll|cc}
\hline
 &  & FID $\downarrow$ & CMMD $\downarrow$\\ \hline
 & Original unrestored LQ images & 53.22 & 0.566  \\ \hline
\multirow{7}{*}{Baselines} & Median Blur & 95.28 & 1.099 \\
 & Deep Image Prior & 57.56 & 0.570 \\
 & Synthetic noise  & 58.24 & 0.356\\
 & CycleGAN & 37.26 & 0.196\\
 & Conditional Diffusion & 47.66 & 0.581 \\
 & CCDF & 43.21 & 0.216 \\
 & Regularized Reverse Diffusion & 38.37 & 0.337 \\ \hline
Ours & RSCD & \textbf{32.02} & \textbf{0.128} \\ \hline
\multirow{6}{*}{Ablations} & No Dynamic Recalibration & 34.62  & 0.137  \\
 & Nonparametric Noise Estimator & 35.53 & 0.270 \\
 & No Step Calibration, 10 steps fixed & 36.41 & 0.174 \\
 & No Step Calibration, 50 steps fixed & 42.06 & 0.224 \\
 & Linear noise schedule & 38.71 & 0.150 \\
 & Unaug. Step Calibrator Training & 33.14 & 0.133 \\\hline
\end{tabular}}
\vspace{-0.1in}
\caption{Results restoring low-quality open-source SRH images from OpenSRH \cite{Jiang2022-cj}. FID score and CMMD score are measured between all restored images and the 4.5K expert-selected high-quality SRH images.}
\label{tab:unpaired}
\vspace{-0.1in}
\end{table}


\cutsubsectionup
\subsection{Evaluation on Near-registered Images}
\label{sec:paired}
\cutsubsectiondown

Next, we aim to evaluate the quality and fidelity of image restoration using an approximately paired low-quality/high-quality SRH imaging dataset. As mentioned before, perfectly paired images are challenging to obtain because optical image degradation is conditional on several (stochastic) factors. However, it is possible to obtain noisy/clean images that are nearly paired, by scanning the same specimen with a cold and warm laser (cold laser is known to produce more noisy images) and then re-align the two images. Through this process, we obtained 2,135 pairs of near-registered low-quality/high-quality SRH patches. These pairs are called `near-registered' because non-affine deformation and optical tissue sectioning make perfect spatial alignment impossible, therefore these images cannot be used as paired training data. Nevertheless, the near-registered SRH dataset is a useful evaluation benchmark for restoration quality and fidelity, as paired data allows additional evaluation metrics (PSNR, SSIM and LPIPS). Details of obtaining near-registered images are in Appendix~\ref{app:near}. 

\begin{figure}
    \centering
    \vspace{-4pt}
    \includegraphics[width=0.48\textwidth]{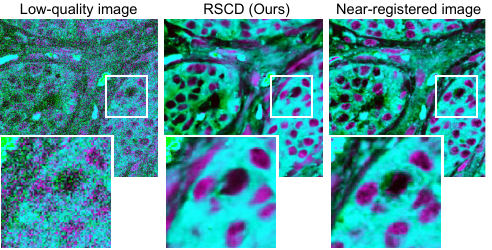}
    
\vspace{-5pt}
    \caption{
    Examples of paired low-quality/near-registered SRH images, and the restored image via RSCD. }
    \label{fig:paired}
\vspace{-10pt}
\end{figure}

\begin{table*}[t]
\centering
\scalebox{0.9}[0.9]{
\begin{tabular}{ll|ccccc}
\hline
   & & FID $\downarrow$ & CMMD $\downarrow$ & PSNR $\uparrow$ & SSIM $\uparrow$ & LPIPS $\downarrow$\\ \hline
  & Paired HQ Images & 0.00 & 0.000 & 100.00 & 1.000  & 0.000 \\
 & Paired LQ Images & 47.64 & 0.506 & 26.97 & 0.604 & 0.598\\ \hline
 \multirow{7}{*}{Baselines} & Median Blur &  72.71 & 1.047 & 28.87 & \underline{0.791} & 0.336 \\
  & Deep Image Prior & 41.51 & 0.385 & \textbf{29.27} & \textbf{0.806} & 0.284 \\
  & Synthetic noise &  77.43& 1.987 & 22.94 & 0.662 & 0.464 \\
  & CycleGAN & 22.03 & 0.185 & 28.39 & 0.779 & 0.667 \\
 
 & Conditional Diffusion & 34.58 & 0.332 & 28.07 & 0.740 & 0.377 \\
 & CCDF & 30.80 & 0.238 & 28.04 & 0.704 & 0.359 \\
& Regularized Reverse Diffusion & 23.37 & 0.270 & 27.89 & 0.682 & 0.459 \\ \hline
 Ours& RSCD & \textbf{21.05} & \textbf{0.104} & \underline{29.06} & \underline{0.791} & \textbf{0.280} \\ \hline
\multirow{6}{*}{Ablations}&  No Dynamic Recalibration & 21.44 & 0.111 & 28.99 & 0.781 & \underline{0.281} \\
& Nonparametric Noise Estimator & 22.77 & 0.215 & 27.90 & 0.691 & 0.418 \\
& No Step Calibrator, 10 steps fixed & 22.58 & 0.163 & 28.29 & 0.721 & 0.357 \\
& No Step Calibrator, 50 steps fixed & 29.72 & 0.136 & 29.05 & 0.788 & 0.335\\
& Linear noise schedule & 32.71 & 0.386 & \underline{29.06} & 0.789 & 0.283 \\
& Unaugmented Step Calibrator Training & \underline{21.15} & \underline{0.105} & 29.03 & 0.786 &  0.284 \\ \hline
\end{tabular}}
\vspace{-0.1in}
\caption{Results of restoring paired low-quality/near-registered high-quality SRH images. FID score is computed between all restored low-quality images and all near-registered images. Bolded numbers are best in each metric, and underlined are second best. RSCD achieved the best FID and LPIPS scores and the second-best PSNR and SSIM across all baselines and ablations.}
\label{tab:pairedbaseline}
\vspace{-0.2in}
\end{table*}

We report image quality and restoration metrics comparing RSCD and above baselines in Table~\ref{tab:pairedbaseline}. Our method achieves the best FID and CMMD, which means our restored images are perceptually the closest to real high-quality images, while most baselines produce blurry or unrealistic images that result in significantly worse FID and CMMD scores. RSCD also has the second highest PSNR and SSIM scores, only behind deep image prior. This indicates that the images restored by our method are close to the ground truth high-quality images while avoiding being excessively smooth and unrealistic (excessively smooth or blurry images, such as the ones produced by median blur or deep image prior, are known to have inflated scores on metrics based on pixel-wise MSE~\cite{wang2009mean}, including PSNR and SSIM). RSCD was the only method to achieve good metrics on both pixel-level and perceptual metrics. 
Examples of near-registered image pairs restored using RSCD are shown in Figure~\ref{fig:paired}, with additional examples shown in Figure~\ref{fig:additionalnearregistered} in Appendix.

%

Our method also achieved top performance among all 5 metrics compared to all ablations, justifying our design choices of using the step calibrator, cosine noise schedule, and dynamic recalibration.


\cutsubsectionup
\subsection{Human Expert Preference}
\cutsubsectiondown

The primary application of RSCD is to restore low-quality SRH images for intraoperative surgical specimen analysis and diagnosis by clinicians and domain experts, so their opinions are \textbf{the most important metric} for evaluating SRH image restoration quality. We recruited three clinicians and three optical imaging experts to evaluate the quality of the restored images and provide preference ratings. Each rater was asked to give their preferences between RSCD restoration and the five baseline methods' restorations. For every preference task, the experts are given the original low-quality image and the two restored images, one from our method and one from the baselines, in random and blinded order. The experts then selected their preferred restored image based on restoration quality and fidelity. In addition, raters were asked if either restoration contained hallucinations. We obtained expert preferences on restorations of 100 randomly selected low-quality images, and we gave every preference task to two raters to measure inter-rater agreement. We include more details about the expert preference collection process in Appendix~\ref{app:humanpref}.

\begin{figure}[t]
    \centering
    \includegraphics[width=0.475\textwidth, bb=0 0 500 450]{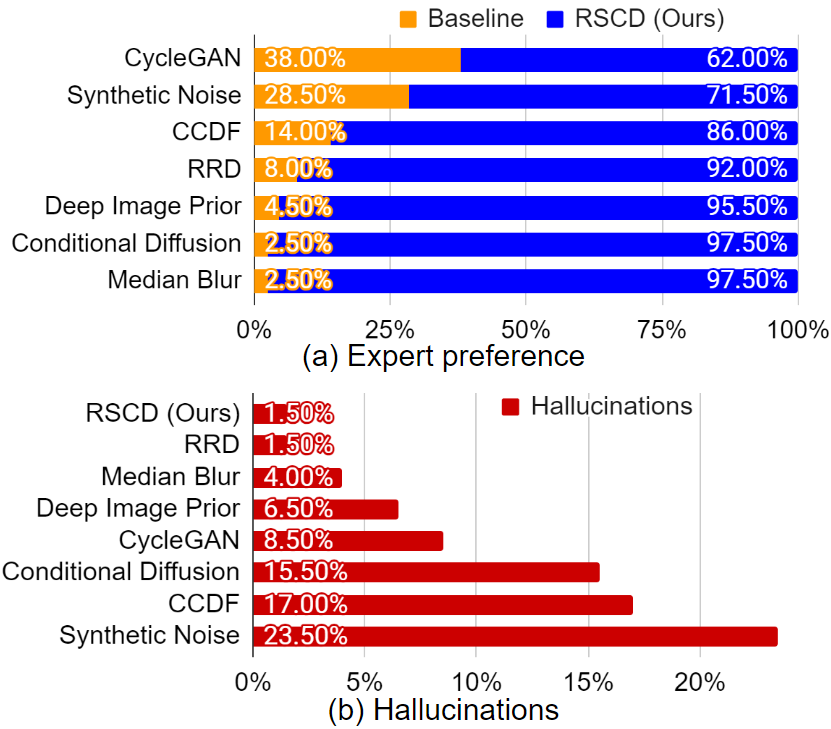}
    \vspace{-17pt}
    \caption{Results of human expert evaluations. \textbf{(a)} The human expert preferences between RSCD and baselines. \textbf{(b)} Percentage of restored images by each method that the experts indicated hallucinated. As shown, the experts overall preferred the restorations from RSCD over the baselines.}
    \vspace{-0.1in}
    \label{fig:human}
\end{figure}

We report the results in Figure~\ref{fig:human}. Human experts preferred RSCD restoration more often than baselines. Raters also reported the least number of hallucinations in our method, less than deterministic methods such as median blur that can create artifacts with severely degraded images. Inter-rater agreement was over 80\% for both quality and hallucination assessments. Our method generates high-quality and reliable restorations of low-quality SRH images according to clinicians and domain experts. 

\cutsectionup
\section{Downstream Clinical Tasks}
\cutsectiondown

\begin{figure}[t]
    \centering
    \includegraphics[width=0.48\textwidth]{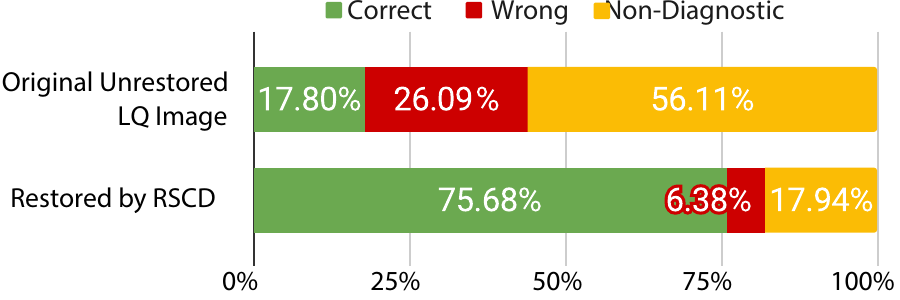}
    
    \vspace{-10pt}
    \caption{Classification accuracy of the SRH tumor classifier from~\cite{Hollon2020-ez} on low-quality images and restored images by RSCD.}\vspace{-10pt}
    \label{fig:classifier}
\end{figure}

In the previous section, we showed that RSCD can restore low-quality SRH images to higher quality, more realistic, and more preferable images compared to other existing unpaired image restoration methods. In this section, we demonstrate the downstream potential of RSCD in restoring SRH images on two clinical tasks: automated deep learning-based tumor diagnostics and z-stack restoration in deep tissue imaging.

\cutsubsectionup
\subsection{Deep learning-based tumor diagnostics}
\label{sec:classifier}
\cutsubsectiondown

Image degradation can decrease the performance of computer vision systems and result in decreased diagnostic accuracy for automated brain tumor classification \cite{Hollon2020-ez}. In this experiment, we test whether RSCD can be used to facilitate more accurate deep learning-based diagnosis of SRH images by restoring model inputs.

We use the SRH tumor classification model from~\cite{Hollon2020-ez}, the most widely accepted study on deep learning-based brain tumor diagnosis using SRH. The model classifies SRH images into one of 3 classes: normal tissue, tumor tissue, or non-diagnostic. The model classifies images as non-diagnostic if the image quality is sufficiently poor that human experts are not able to determine the underlying tissue diagnosis. For each unpaired low-quality image, we pass both the original low-quality SRH image and the restored image through the SRH tumor classification model. We report the classification results in Figure~\ref{fig:classifier}.

Without restoration, over 50\% of the SRH images were classified as non-diagnostic and another 26\% were incorrectly classified. After restoration with RSCD, the classifier correctly diagnosed over 75\% low-quality SRH images, and only miss-classified 6\% of images. Restoring SRH images via RSCD can significantly improve deep learning-based automated diagnostic accuracy, and can drastically reduce non-diagnostic predictions caused by low image quality. Importantly, SRH restoration significantly reduces the risk of a wrong diagnosis, which can have a severe detrimental effect on patient care and surgical treatment. RSCD can make existing automated diagnostic tools safer and more reliable. We include more details and examples in Appendix~\ref{app:diagnosis}.


\cutsubsectionup
\subsection{Z-stack Image Restoration}
\cutsubsectiondown

A known limitation of Raman-based optical imaging is that signal-to-noise ratios decrease as imaging depth increases due to laser scattering and absorption by tissue above the scanned depth. In this experiment, we show that RSCD can restore z-stack data, which contains SRH images acquired at sequentially deeper spatial locations in the tissue. z-stack images are volumetric and capture 3-D biological structures by imaging in all three spatial dimensions.

\begin{figure}[t]
    \centering
    \includegraphics[width=\columnwidth]{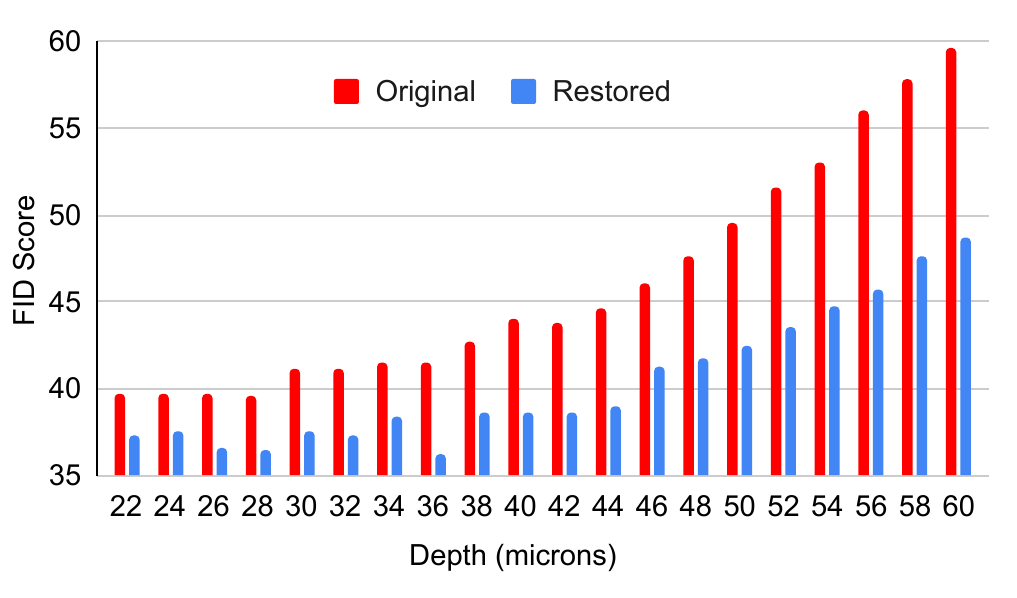}
    \vspace{-15pt}
    
    \caption{FID scores of original and restored z-stack images. Lower FID scores indicate better perceptual quality. RSCD consistently improves image quality at all depths.}
    \vspace{-0.15in}
    \label{fig:zstack}
\end{figure}

RSCD was used to restore z-stacked SRH images taken from 425 surgical specimens. Additional information regarding the dataset can be found in Appendix~\ref{app:zstack}. FID scores were computed at each z-depth level for original and restored images with respect to the 4.5K high-quality SRH images. We show the results in Figure~\ref{fig:zstack}. FID scores for the original low-quality images steadily increase as imaging depth increases, indicating worse image quality with imaging depth. RSCD consistently reduces the FID score for deep SRH images, as shown in Figure~\ref{fig:depth}. We present additional examples of z-stack restoration by RSCD in  \ref{fig:zstackadditional} in Appendix~\ref{app:zstack}.

\begin{figure*}[t]
    \centering
    \vspace{-5pt}
    \includegraphics[width=\textwidth]{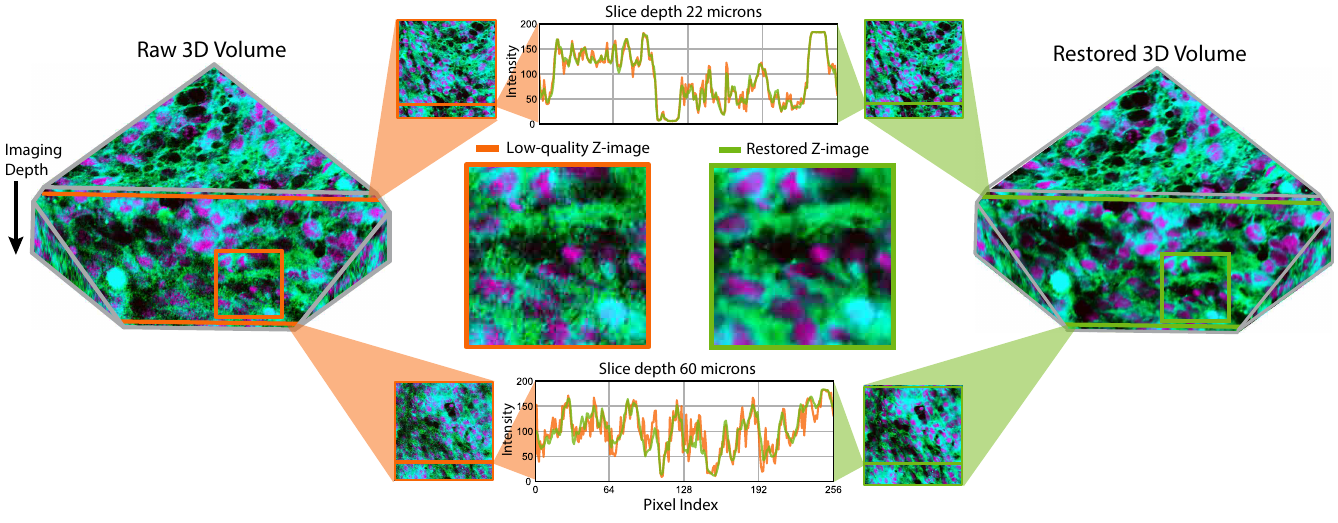}
    \vspace{-5pt}
    \caption{Depth z-stack restoration results. The raw 3D z-stack volume (left) and the restored volume (right) are shown. Depth z-images in the center show image restoration using RSCD. The top and bottom plots show the pixel intensities along the row indicated in orange and green in the adjacent images. As depth increases, image noise increases as shown in the orange, low-quality pixel intensities. However, images restored by RSCD show little to no additional noise with increasing depth.}
    \label{fig:depth}
\end{figure*}

\section{Related works}
\label{sec:relwork}

Several previous works share the similarity of using only the last part of the reverse diffusion generative process for image generation or restoration purposes (instead of requiring the full reverse diffusion) with our work. TDPM~\cite{zheng2022truncated} increases the efficiency of diffusion-based image generation by replacing the majority of the reverse diffusion steps with a GAN and only performing the last part of the reverse diffusion process. SDEdit~\cite{meng2021sdedit} and CCDF~\cite{chung2022come} perform image editing and restoration by first adding a strong Gaussian noise to the input image, then performing a few reverse diffusion steps. Regularized Reverse Diffusion (RRD)~\cite{Chung2023-qt} builds upon CCDF: it does not require noise addition, uses a non-parametric Gaussian noise level estimator~\cite{Chen2015-ij} to determine the number of reverse diffusion steps to perform, and regularizes the restoration process with Fourier features.  In comparison, RSCD has the following advantages: (1) we designed and trained a highly effective step calibrator to determine the number of steps required, while TDPM, SDEdit, and CCDF set this as a fixed hyperparameter and RRD determines this with a non-parametric Gaussian noise level estimator, all of which does not work well when the input image has non-uniform noise of varying strengths (as evident by our ablation studies on fixed steps and nonparametric noise estimator);  (2) Adding a large amount of noise at once to the input image (like SDEdit and CCDF) increases the risk of hallucination, which is unacceptable for medical images; instead, RSCD does not add a large amount of noise at once; RSCD mitigates the noise distribution difference between real noise and Gaussian noise by gradually adding back a small amount of Gaussian noise during each DDPM step and dynamic recalibration.

We include a more comprehensive discussion of related works on image restoration and deep learning applications in biomedical optical imaging in Appendix~\ref{app:rel}.


\section{Limitations}
\label{app:lim}

Our work is motivated by a biomedical imaging problem and intentionally focuses on biomedical optical imaging in the clinical setting. Our strategy is motivated by the problem of restoring medical images corrupted through non-reproducible, stochastic degradation sources and must be resistant to hallucinations. This problem is not limited to biomedical optical imaging, so additional studies are needed to confirm whether RSCD generalizes to other settings as additional related datasets are released. Another limitation is that the current version of RSCD does not take into account neighboring spatial regions when restoring a specific field of view. SRH image patches are sampled from megapixel images and consistent restoration should be enforced across the full image. We are currently exploring strategies to incorporate neighboring patches into the restoration process to improve restoration consistency, such as sliding windows attention \cite{Esser2020-kv} or unified conditional diffusion~\cite{Zhang2024-bc}.

\section{Conclusion}
\label{sec:conclusion}

We present Restorative Step-Calibrated Diffusion (RSCD), a reliable and efficient method to restore biomedical optical imaging without requiring paired high-quality data. RSCD outperforms other widely used unpaired image restoration methods on quantitative metrics; more importantly, experts in biomedical optical imaging consistently prefer images restored using RSCD in blinded comparison experiments and report minimal to no hallucinations. RSCD can improve model performance on downstream clinical tasks, including automated brain tumor diagnosis and deep tissue imaging. Our method reduces diagnostic errors that can have detrimental impacts on clinical care. This study demonstrates the potential of AI in improving automated clinical diagnostics and patient care in today's precision medicine landscape.

\section*{Data Usage Ethics Statement}

The SRH datasets used in this study include specimens from patients who underwent brain tumor biopsy or tumor resection. Patients were consecutively and prospectively enrolled
at University of Michigan for intraoperative SRH imaging, and this study was approved by the Institutional Review Board (HUM\#00083059). Informed consent was obtained for each patient prior to SRH imaging and
the use of tumor specimens for research and development was approved.

\section*{Acknowledgements}

We would like to thank Karen Eddy, Lin Wang, Hubert Zhang for their administrative support and data
collection.

This work was supported by the following National Institute of Health (NIH) funding sources: K12NS080223
(T.C.H.), T32GM141746 (C.J.), F31NS135973 (C.J.). This work was support by the Chan Zuckerberg
Foundation (CZI) Advancing Imaging Through Collaborative Project grant, Cook Family Brain Tumor
Research Fund (T.C.H), the Mark Trauner Brain Research Fund, the Zenkel Family Foundation (T.C.H.),
Ian’s Friends Foundation (T.C.H.) and the UM Precision Health Investigators Awards grant program
(T.C.H.).

This research was also supported, in part, through computational resources and services provided by
Advanced Research Computing, a division of Information and Technology Services at the University of
Michigan.

T.C.H. and C.F. are shareholders of Invenio Imaging, Inc., a company developing SRH microscopy systems.

\bibliography{11_references,paperpile}

\clearpage

\setcounter{secnumdepth}{2}

\appendix
\label{sec:appendix}


\section{Additional Related Works}
\label{app:rel}

\cutsubsectionup
\subsection{Image Denoising and Restoration}
\cutsubsectiondown

Image denoising and restoration belong to a set of problems called image inverse problems~\cite{Ongie2020-zi}. There are several different levels of difficulty for image denoising and restoration problems. The easiest level is when the corruption process is known as a linear process (such as blurring, inpainting, uniform Gaussian noise, etc) and can be programmatically replicated onto clean images, and the goal is to create an inverse process to the corruption. There have been extensive studies of restoring known corruptions, including supervised approaches~\cite{Ongie2020-zi} and unsupervised approaches~\cite{Kawar2022-vc,Kadkhodaie2021-wd,wang2022zero}. The second level of difficulty is when the corruption process is not reproducible, but there are existing pairs of clean and corrupted images allowing for supervised training, and the typical solution is training models to generate the clean image given the corrupted images in a supervised way, such as UNet~\cite{Ronneberger2015-um,Manifold2019-ft} or Conditional Diffusion~\cite{Choi2021-mx,Kawar2022-vc, Saharia2022-co}. The most difficult setting is when we don't have paired data and also cannot reproduce the corruption programmatically; instead, we only have access to unpaired clean images and corrupted images. Existing approaches for this challenging setting include cycle-consistency-based approaches such as CycleGAN~\cite{Zhu2017-wf} and UnitDDPM~\cite{Sasaki2021-ks}, VAE-based approaches~\cite{Zheng2023-dz}, and flow-based approaches~\cite{Du2020-om}. In addition to the above settings, there are also denoising methods that do not require a training dataset, such as MedianBlur or Deep Image Prior~\cite{Ulyanov2016-kz}. General Diffusion Prior~\cite{fei2023generative} is a recent unsupervised diffusion-based method that can perform unpaired non-linear blind image restoration, but it is extremely computationally costly as it requires full generative diffusion with multiple gradient steps within each reverse diffusion step.

Restoring noisy SRH images generally falls within the third difficulty level (i.e. no paired data and not a linear inverse problem), as the distribution of noise is unpredictable. Although it is possible to apply supervised methods from the second level by training to remove Gaussian and Laser noises (such as our UNet and Conditional Diffusion baselines), they are generally outperformed by unpaired methods as shown in our experiments. Several recent approaches (including~\cite{chung2022come},~\cite{Chung2023-qt}, and~\cite{he2023tdiffde}) also used the last part of the reverse diffusion process for unpaired image restoration. We discussed these closely related works in more detail in Section~\ref{sec:relwork}.

\cutsubsectionup
\subsection{Deep Learning applications on SRH}
\cutsubsectiondown

Deep learning techniques have been playing an important role ever since SRH was used for intra-operative neurosurgery, as the very first paper that introduces this optical diagnostic technique~\cite{Orringer2017-nn} trained a multi-layer perceptron to automatically predicts brain tumor subtypes from quantified SRH image attributes. Later, Hollon et al.~\cite{Hollon2020-ez} proposed a CNN-based model that can directly interpret raw SRH images (instead of quantified attributes) and make diagnosis at over 94\% accuracy. The first public dataset of clinical SRH images, OpenSRH~\cite{Jiang2022-cj}, was released in 2022, and it contains SRH images from 300+ brain tumor patients and 1300+ unique whole slide optical images. DeepGlioma~\cite{Hollon2023-uf} has allowed deep-learning-based models to perform molecular classification of diffuse glioma types via a multimodal model that takes in both SRH images and diffuse glioma genomic data. HiDisc~\cite{Jiang2023-rq} and S3L~\cite{hou2024selfsupervised} are self-supervised visual representation learning methods that were developed to classify whole-slide SRH images.

All the above works focused on the classification of SRH images, and there have been few previous works that attempted to tackle the unavoidable noise problem of SRH images through image restoration. To the best of our knowledge, the only previous work on this topic was~\cite{Manifold2019-ft}, in which the authors collected a paired dataset using low and high laser energy, and trained a U-Net to perform the denoising. In comparison, our method works for all types of noisy SRH images (not just those resulting from low laser power), and we have shown that our method outperforms their U-Net approach.

\section{Details about SRH images}

\begin{figure*}
    \centering
    \includegraphics[width=\textwidth,bb=0 0 601 598]{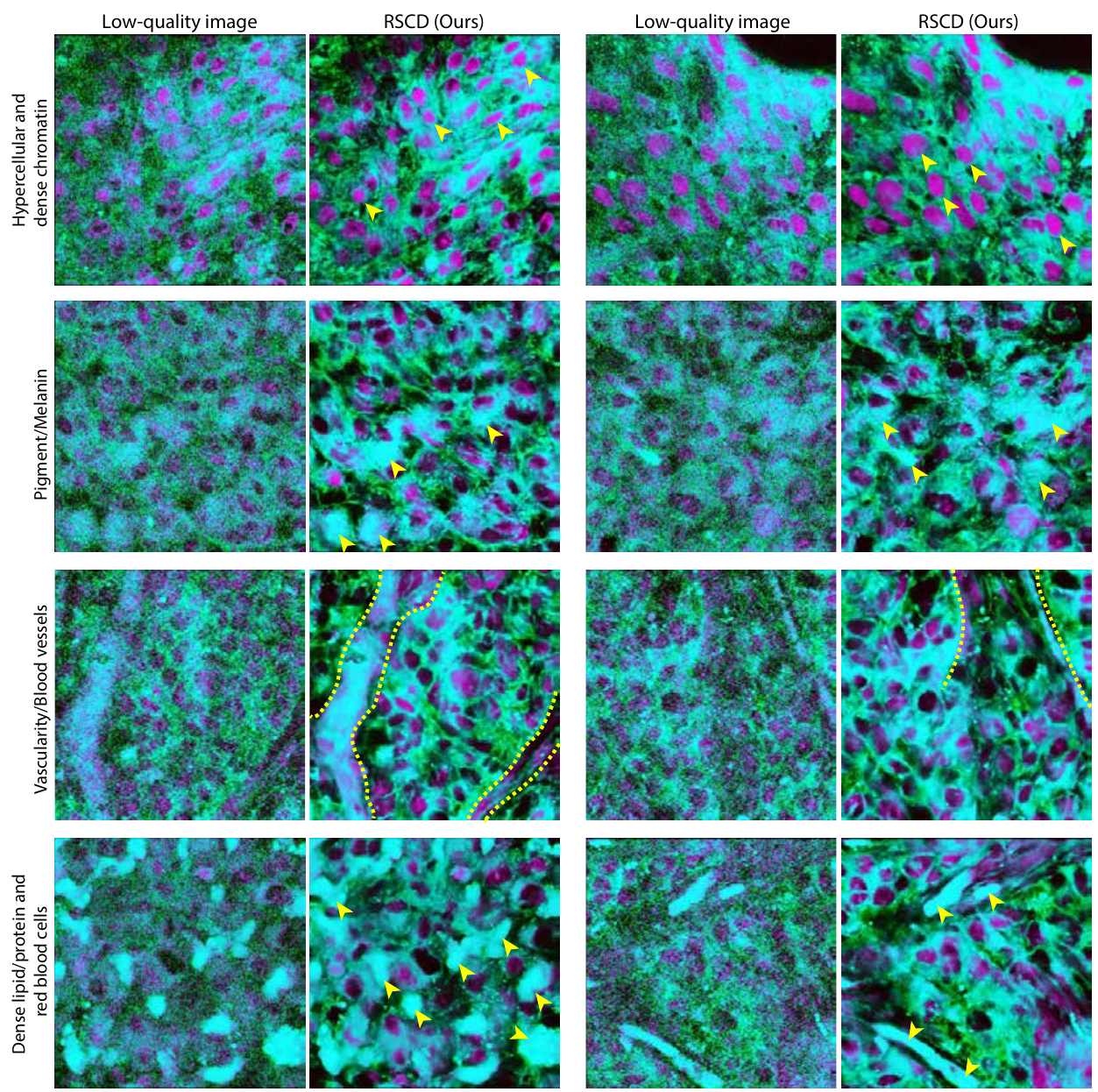}
    \caption{Examples of tissue-dependent sources of image degradation. Each row is an example of cytologic or histologic tissue features that can result in excessive laser scattering, absorption, and image degradation. \textbf{First row:} hypercellularity and dense chromatin (yellow arrows) increase the overall density of tissues. \textbf{Second row:} melanin and other pigments (yellow arrows) can have autofluorescence and reduce signal-to-noise ratios. Third row, blood vessels (yellow borders) contain collagen and elastin which are dense, proteinaceous structures that scatter laser. \textbf{Last row:} the same is true for lipid- or protein-dense intracellular inclusions and red blood cells (yellow arrows).}
    \label{fig:tumornoiseadditional}
\end{figure*}

We include examples of image degradation for different cytologic/histologic features and brain tumor types in Figure~\ref{fig:tumornoiseadditional}. The examples show there exist many vastly different types of degradation that can randomly occur in low-quality SRH images, therefore making SRH image restoration a challenging task, but RSCD was able to restore them quite well.

\section{Methodology Additional Details}

\subsection{Automated Data Filtering}
\label{app:auto}

Similar to large language model data curation strategies \cite{Brown2020-dv}, we used a classifier-based filter method to obtain the full high-quality diffusion model training set. A binary classifier was trained on the 4.5K annotated high-quality patches as positive examples (negative examples were randomly selected from the remaining uncurated SRH patch dataset). The trained classifier was then used to split the full SRH dataset into low-quality/high-quality images, resulting in a total of $\sim$840K high-quality SRH images for training.

\subsection{Step Calibrator Training and Calibration Distribution}
\label{app:steppred}

\begin{figure}[b]
     \centering
     \includegraphics[width=1.0\columnwidth,bb=0 0 950 450]{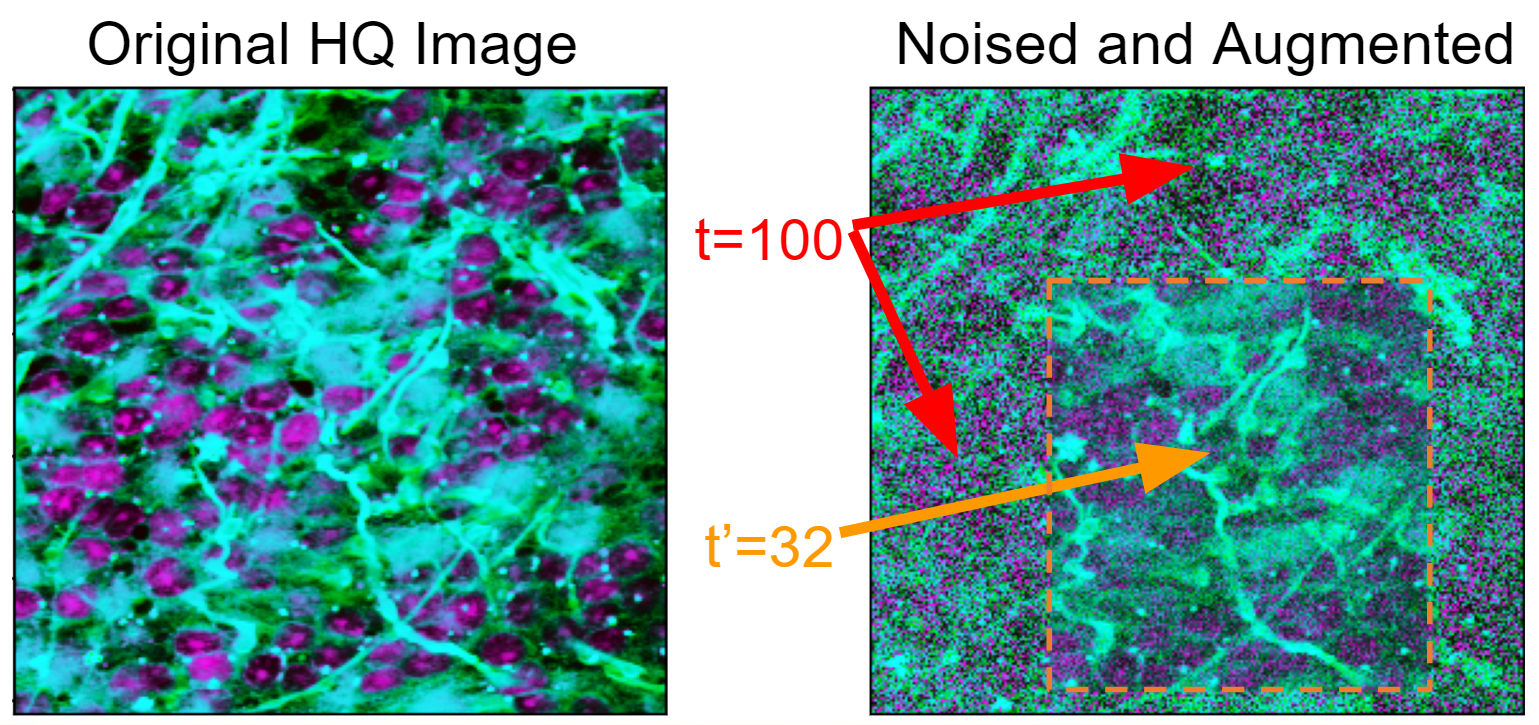}
     \caption{An example of the noise augmentation applied for training better step calibrator models. In this example, we can see that the augmented image is more noisy around the edges ($t=100$) while is less noisy around the center ($t'=32$).}
     \label{fig:aug}
\end{figure}

We show the full algorithm of training the step calibrator with manually added Gaussian noise and augmentations in Algorithm~\ref{alg:sp}, as well as one example of the augmentation in Figure~\ref{fig:aug}.

\begin{algorithm}
\caption{Training step calibrator with \textcolor{red}{augmentations}}\label{alg:sp}
\begin{algorithmic}[1]
\State \textbf{Requires:} Set of high quality images $X$, total steps $T$, hyperparameters based on noise schedule $\bar{\alpha}_1$,...,$\bar{\alpha}_T$ 
\State $S \gets$ ResNet-50
\Repeat
\State Sample $x \in X$
\State $t\sim$ Uniform(\{1,...,T\})
\State $\epsilon\sim N(0,1)$
\State $x_t \gets \sqrt{\bar{\alpha}_t}x+\sqrt{1-\bar{\alpha}_t}\epsilon$ 
\color{red}
\State $t' \sim$ Uniform(\{1,...,t\})
\State $\epsilon'\sim N(0,1)$
\State $x_{t'} \gets \sqrt{\bar{\alpha}_{t'}}x+\sqrt{1-\bar{\alpha}_{t'}}\epsilon'$ 
\State $m \gets $ Binary mask of random region
\State $x_t = mx_t+(1-m)x_{t'}$
\color{black}
\State $L \gets ||t-S(x_t)||_2$
\State Train $S$ with loss $L$
\Until{converged}
\end{algorithmic}
\end{algorithm}

\begin{figure}[]
     \centering
     \includegraphics[width=1.0\columnwidth,bb=0 0 383 95]{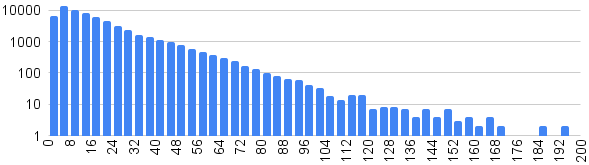}
     \caption{Distribution of the initial step calibrations for low-quality SRH image restoration. The number of images decreases exponentially as the step number increases.}
     \label{fig:histogram}
\end{figure}

To determine a proper $T'$, we ran our trained step calibrator on 63K real low-quality SRH images, and the distribution of $t_{pred}$ is shown in Figure~\ref{fig:histogram}. We can see that the number of images decreases exponentially with its $t_{pred}$, and $t_{pred}$ never exceeded 200, which justifies our choice of $T'=200$ in our diffusion model training.

\subsection{Diffusion Model Training}
\label{app:train}

We show the full training algorithm in Algorithm~\ref{alg:train} below. It is almost identical to the training algorithm from~\cite{Ho2020-tm}, except that we only sample $t$ up to $T'=200$.

\begin{algorithm}
\caption{Restorative Step-Calibrated Diffusion: Training with \textcolor{red}{shortcut}}\label{alg:train}
\begin{algorithmic}[1]
\State \textbf{Requires:} Set of high-quality images $X$, diffusion model $\epsilon_\theta$, total steps $T$, \textcolor{red}{max training steps $T'$}, noise schedule hyperparameters $\bar{\alpha}_1$,...,$\bar{\alpha}_T$ 
\Repeat
\State Sample $x_0 \in X$
\State $t\sim$ Uniform(\{1,...,\textcolor{red}{$T'$}\})
\State $\epsilon\sim N(0,1)$
\State Take gradient descent step on \\
    \; \; \; \; \; $\nabla_\theta ||\epsilon - \epsilon_\theta(\sqrt{\bar{\alpha}_t}x_0+\sqrt{1-\bar{\alpha}_t}\epsilon,t)||^2$
\Until{converged}
\end{algorithmic}
\end{algorithm}

\section{Experiment details and additional results}
\label{app:experiments}

Due to the scale of the experiments, all methods are only run once with one fixed random seed across all images.

\subsection{Baseline and Ablation details}
\label{app:baseline}

Below are the modeling and training details of the baselines:

\textbf{CycleGAN \cite{Zhu2017-wf}:} We trained a CycleGAN model for image-to-image translation between unpaired low-quality and high-quality images. We used the 4.5K hand-picked high-quality and 3.8K hand-picked low-quality images for training the CyleGAN. Both the high-quality and low-quality image sets contained at least 10 different tumor types. The CycleGAN model performed better when trained on the manually selected images compared to training on the full 840K filtered high-quality images.

\textbf{Synthetic noise\cite{Lehtinen2018-zd, Manifold2019-ft}:} A U-Net model is trained by adding Gaussian noise to high-quality images to generate synthetic low-quality images. We train only with the 4.5K manually selected high-quality images because we found that the model performs better compared to training on the full 840K filtered high-quality images.

\textbf{Conditional Diffusion~\cite{Kawar2022-vc, Saharia2022-co}:} We train a conditional generative diffusion model, where we sample from a random Gaussian prior, and generate the restored image conditioned on the low-quality image. The conditioning method is concatenation (i.e. the input image $x_t$ and the conditioning image is concatenated as input to the UNet). We train the model with the 840K HQ training images and fine-tune it with the 4.5K hand-picked images. During training, we add a random amount of Gaussian noise to the high-quality images as conditioning images. We show the full training algorithm in Algorithm.

\begin{algorithm}
\caption{Conditional Diffusion: Training}\label{alg:condtrain}
\begin{algorithmic}[1]
\State \textbf{Requires:} Set of high-quality images $X$, diffusion model $\epsilon_\theta$, total steps $T$, max training steps $T'$, noise schedule hyperparameters $\bar{\alpha}_1$,...,$\bar{\alpha}_T$ 
\Repeat
\State Sample $x_0 \in X$
\State $t'\sim$ Uniform(\{1,..., T'\})
\State $\epsilon'\sim N(0,1)$
\State $x_{cond} \gets \sqrt{\bar{\alpha}_{t'}}x_0+\sqrt{1-\bar{\alpha}_{t'}}\epsilon$
\State $t\sim$ Uniform(\{1,..., T\})
\State $\epsilon\sim N(0,1)$
\State Take gradient descent step on \\
    \; \; \; \; \; $\nabla_\theta ||\epsilon - \epsilon_\theta(\sqrt{\bar{\alpha}_t}x_0+\sqrt{1-\bar{\alpha}_t}\epsilon,t,x_{cond})||^2$
\Until{converged}
\end{algorithmic}
\end{algorithm}

\textbf{Regularized Reverse Diffusion (RRD) \cite{Chung2023-qt}} is a recent method developed for restoring noisy medical images. It also uses part of the reverse diffusion process as restoration, but it employs a non-parametric noise estimator \cite{Chen2015-ij} to determine the step numbers, and uses a Fourier-feature-based regularization to reduce hallucination. Since we were unable to find any official code for the RRD paper and we deal with images in a vastly different domain from theirs (RRD was originally designed for denoising single-channel MRI images), we trained our own diffusion model following the same process as~\cite{Ho2020-tm} and using the same cosine noise schedule as RSCD.

\textbf{CCDF~\cite{chung2022come}} is a recent method for efficient diffusion-based image editing and denoising. CCDF first adds a certain number of steps of Gaussian noise to the input image, and then perform reverse diffusion to create the output image. The number of steps is a fixed hyperparameter, and we follow the original CCDF paper to pick a fixed step number of 20. We trained our own diffusion model following the same process as \cite{Ho2020-tm} and using the same cosine noise schedule as RSCD.

\textbf{Deep Image Prior~\cite{Ulyanov2016-kz}} is an unsupervised image restoration method that works by training a ResNet to reconstruct the noisy image but takes the generation of the model before its convergence as the denoised image. We take the output of step 1800 as our baseline, as images generated from this training step tend to have the best quality.

\textbf{Median Blur} is a traditional, but effective, method to smooth noisy images. It replaces each pixel value with the median of the $k \times k$ pixel kernel ($k=5$ for experiments). 

Below are details of the ablations:

\textbf{No Dynamic Recalibration}: RSCD without Dynamic Recalibration, i.e. Algorithm~\ref{alg:dr} without the blue parts. We take the initial step calibration and directly run that many steps of denoising to the end.

\textbf{Nonparametric Noise Estimator}: We replace our ResNet-based step calibrator with the nonparametric noise estimator~\cite{Chen2015-ij} used in RRD~\cite{Chung2023-qt}. We determine the step number $t$ as the closest cumulative noise $\sqrt{1-\bar{\alpha}_t}$ to the predicted noise level from the nonparametric noise estimator.

\textbf{No Step Calibrator, 10 steps always}: Instead of using a step calibrator and Dynamic Recalibration, we always only perform 10 steps of diffusion-based denoising on the low-quality images.

\textbf{No Step Calibrator, 50 steps always}: Instead of using a step calibrator and Dynamic Recalibration, we always only perform 50 steps of diffusion-based denoising on the low-quality images.

\textbf{Linear noise schedule}: We use a linear noise schedule instead of the cosine linear schedule.

\textbf{No Augmentation Step Calibrator}: We use a step calibrator that is trained without augmentations (i.e. Algorithm~\ref{alg:sp} without the red parts).

\subsection{Unpaired Images Evaluation additional qualitative results}

\begin{figure*}
    \centering
    \includegraphics[width=\textwidth,bb=0 0 702 599]{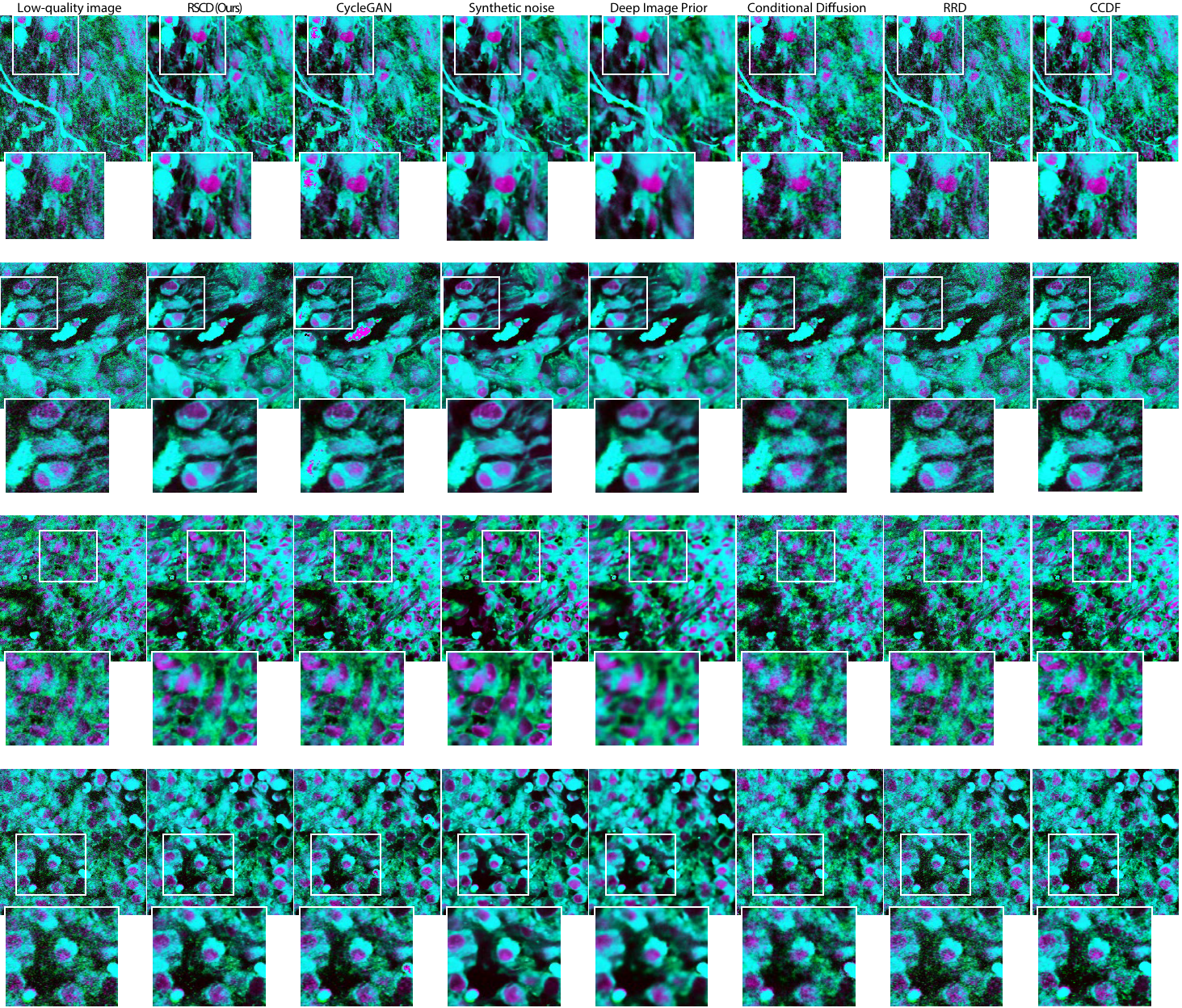}
    \caption{Additional qualitative results for the unpaired experiment.}
    \label{fig:unpairedadditional}
\end{figure*}

We show additional qualitative comparisons of restorations of unpaired images in Figure~\ref{fig:unpairedadditional}.

\subsection{Near-registered data collection details}
\label{app:near}

Imaging with low laser power is one of the known ways to produce noisy images. We generate a paired set of near-registered SRH images by imaging a surgical specimen immediately after the SRH imaging system is started (without laser warm-up and optimization), and then image the same specimen after 20 minutes, once the laser has been optimized and sufficiently warmed up. Next, image registration is required to spatially align the low-quality and high-quality images because the tissue specimen may be shifted due to gravity between the two image acquisitions. Paired images were registered by fitting an affine transformation using scale-invariant feature transform \cite{Lowe2004-gq} and RANSAC~\cite{derpanis2010overview}.

While the near-registered images are matched enough for meaningful evaluation of fidelity, they are not fitting for training supervised image restoration because (1) the images are not perfectly paired (free-floating structures like red blood cells may have rotated/shifted relative to nearby cells), so training with them is just promoting hallucinations; (2) the noise in this dataset results from cold laser alone, so it does not well represent all types of random noise that could occur in SRH images from real intraoperative settings.

\subsection{Near-registered evaluation additional examples}

\begin{figure*}
    \includegraphics[width=\textwidth]{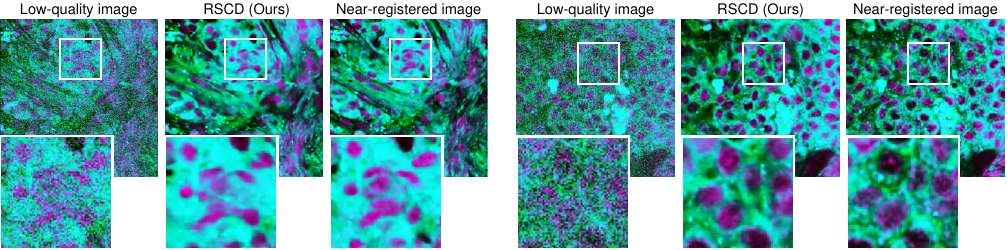}
    \caption{Additional comparison examples for the experiment with near-registered images.}
    \label{fig:additionalnearregistered}
\end{figure*}

We show additional examples of comparisons between the original/restored/near-registered images from the near-registered experiment in Figure~\ref{fig:additionalnearregistered}.

\begin{table}[]
\centering
\begin{tabular}{l|cc}
\hline
 & FID $\downarrow$ & CMMD $\downarrow$  \\ \hline
$d=10$ (ours) & 32.02 & 0.128 \\
$d=5$ & 32.30 & 0.132 \\
$d=20$ & 32.64 & 0.126 \\
\hline
\end{tabular}
\caption{Results of ablation study on different recalibration intervals ($d$)}
\label{tab:additionalablationd}
\end{table}

\subsection{Hyperparameters}
\label{app:hyperparameter}

We list all architecture/training/inference hyperparameters for all components of RSCD in Table~\ref{tab:hyperparameter}. The number of training epochs are determined based on convergence of loss. Learning rate was chosen between $1e-3$ to $1e-6$ for optimal training efficiency and performance. Batch size was limited by GPU memory.

\begin{table*}[]
\centering
\begin{tabular}{l|l|l}
\hline
\multirow{6}{*}{\begin{tabular}[c]{@{}l@{}}Step\\ Calibrator\\ Training\end{tabular}} & Base Model & ResNet-50 \\ \cline{2-3}
 & Prediction Head & \begin{tabular}[c]{@{}l@{}}2-Layer MLP \\(2048,100,1)\\ ReLU activation\end{tabular} \\\cline{2-3}
 & Learning Rate & 0.00001 \\\cline{2-3}
 & Optimizer & Adam \\\cline{2-3}
 & Epochs & 500 \\ \cline{2-3}
 & Batch Size & 150 \\\hline
\multirow{7}{*}{\begin{tabular}[c]{@{}l@{}}Diffusion\\ Model\\ Architecture\end{tabular}} & Model & UNet \\\cline{2-3}
 & In channels & 2 \\\cline{2-3}
 & Init dim & 128 \\\cline{2-3}
 & Out channels & 2 \\\cline{2-3}
 & Dim mult & {[}1,2,2,4{]} \\\cline{2-3}
 & Resnet block groups & 8 \\\cline{2-3}
 & Convnext Mult & 2 \\ \hline
\multirow{8}{*}{\begin{tabular}[c]{@{}l@{}}Diffusion\\ Model\\ Training\end{tabular}} & Noise Schedule & Cosine (s=0.008) \\\cline{2-3}
 & Total Steps (T) & 1000 \\\cline{2-3}
 & \begin{tabular}[c]{@{}l@{}}Max Shortcut Training\\ Steps (T')\end{tabular} & 200 \\\cline{2-3}
 & Batch Size & 8 \\\cline{2-3}
 & Learning Rate & 0.00001 \\\cline{2-3}
 & Loss type & L1 \\\cline{2-3}
 & Epochs & \begin{tabular}[c]{@{}l@{}}1 on 840K (full data)\\ 20 on 4.5K (finetuning)\end{tabular} \\\cline{2-3}
 & Optimizer & Adam \\ \hline
\multirow{2}{*}{\begin{tabular}[c]{@{}l@{}}Dynamic\\ Recalibration\end{tabular}} & \multirow{2}{*}{Recalibration interval (d)} & \multirow{2}{*}{10} \\ & & \\
  \hline
\end{tabular}
\caption{Modeling, training and inference hyperparameters for all components of RSCD}
\label{tab:hyperparameter}
\end{table*}

\subsection{Computing Resource and Inference Efficiency}
\label{app:compres}

Training the step calibrator takes about 5 hours on 1 A40 GPU. Training the diffusion model on 840K data took about 4 days on 1 A40 GPU, and fine-tuning on 4.5K hand-picked data took about 1 day on 1 A40 GPU. 

We analyze the inference efficiency of RSCD on a single RTX 2080 GPU, in both runtime and NFE of the forward function of the UNet. On average over 1000 randomly selected low-quality SRH images, RSCD takes about 14.5 seconds to denoise a 256x256 patch (34 NFE), while methods that runs full generative reverse diffusion (e.g. Conditional Diffusion) take 431 seconds per patch (1000 NFE). Thus, RSCD is about 30x faster than running a full generative reverse diffusion process. Also, note that this number is obtained by denoising each patch individually without batching on an old and slow GPU for fair comparison. The amortized computation time can be improved by a lot if we perform the denoising steps in large batches and with a better GPU.

\subsection{Human Expert Preference Details}
\label{app:humanpref}

\begin{figure*}
    \centering
    \includegraphics[width=\textwidth]{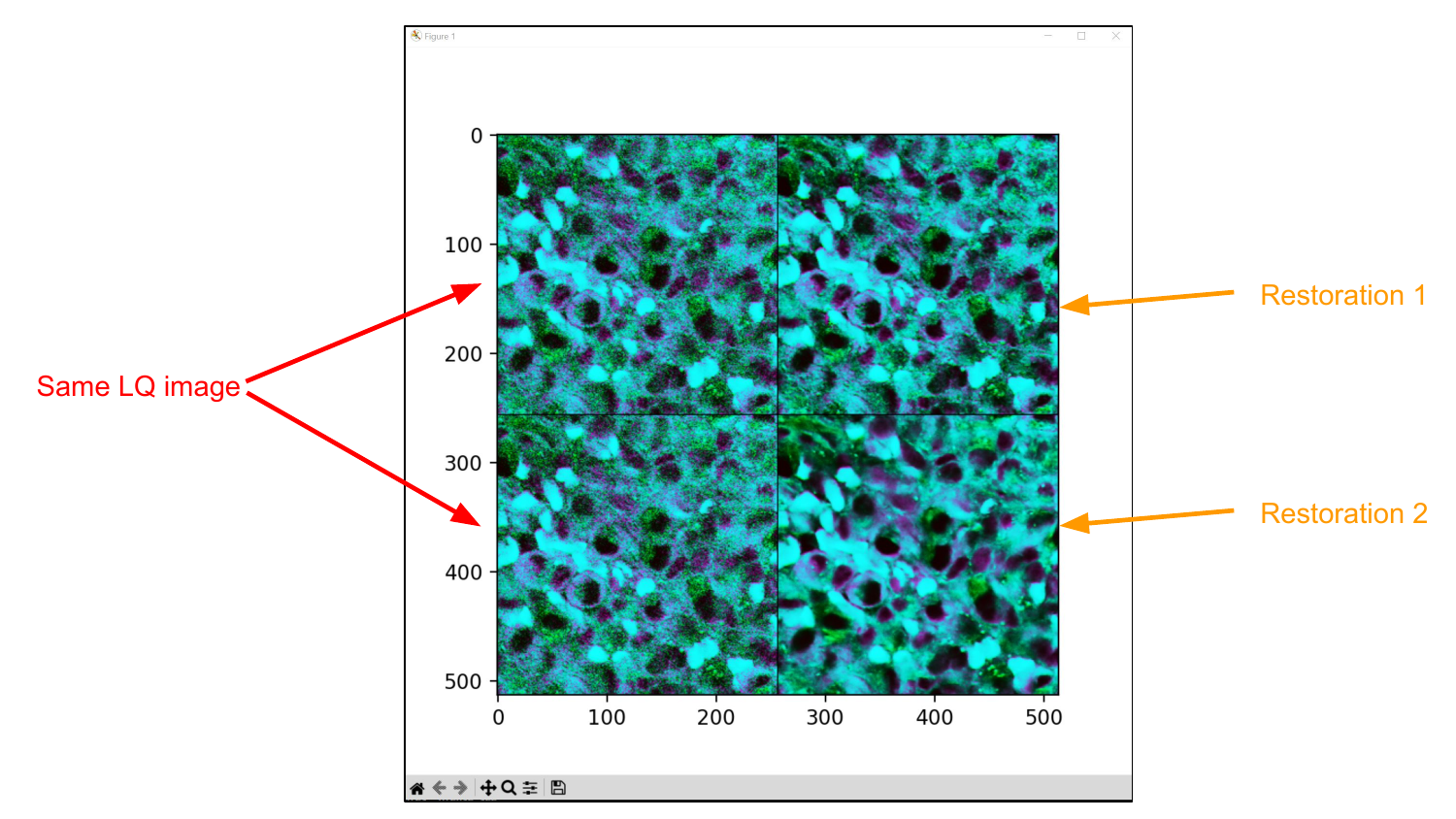}
    \caption{The interface used for human expert preference. The expert is presented with the original LQ image on the left and 2 different restorations on the right. The expert is then asked to indicate their preference on which restoration has better overall quality and which ones have clinically significant hallucinations.}
    \label{fig:interface}
\end{figure*}

We show the human expert preference collection interface in Figure~\ref{fig:interface}. The expert is given the original LQ image as well as 2 restorations (one from RSCD and one from a baseline, in random order; the expert is not told which restoration is baseline and which one is RSCD), and then the expert is asked to (1) select one restoration that the expert believes has better overall quality, and (2) determine which restoration has clinically significant hallucinations (can choose "neither" if neither restoration hallucinates). We perform a total of 1200 such pair-wise preference collections (100 LQ images x 6 baseline comparisons x 2 preferences from different experts per comparison), distributed among all participating experts.

\subsection{Additional Ablation on Recalibration Interval}

We performed additional ablation studies to explore the effect of different recalibration intervals ($d$) on the unpaired low-quality images. The results are shown in Table~\ref{tab:additionalablationd}. We found that when we change $d$ from 10 to 5 or 20, the difference in performance of model is negligible compared to differences between RSCD and other baselines/ablations in Table~\ref{tab:unpaired}.

\section{Downstream task details}

\subsection{Deep learning-based tumor diagnosis}
\label{app:diagnosis}

\paragraph{Data:} A portion of our SRH images have ground truth labels on whether the tissue is normal or tumor. Therefore, we took the intersection of that labeled portion and the 12K unpaired low-quality images used for the experiment in section~\ref{sec:unpaired} as the dataset for deep learning-based tumor diagnosis. The dataset contains 8,372 low-quality images, each with its ground truth label of tumor/normal.

\paragraph{Model:} We used the segmentation model from~\cite{Hollon2020-ez}, the most influential and widely accepted study on deep learning-based brain tumor diagnosis through SRH images, as our classification model. The model classifies each SRH image into one of three classes: normal, tumor, and non-diagnostic. The model was trained to classify an SRH image as non-diagnostic if the model is unable to extract sufficient information from the image to make a diagnosis, and there are several possibilities for an image to be non-diagnostic, including: (1) the image scanning an irrelevant part of the slide, or (2) the image quality is too low.

\paragraph{Evaluation:} For each image in our dataset, we run the same classification model on both the original and the restored version of the image. We categorize the predictions into 3 categories as follows:

\begin{itemize}
    \item \textbf{Correct: } If the ground truth label matches the predicted label (i.e. predicting an image from a tumor tissue as tumor, or predicting an image from a normal tissue as normal)
    \item \textbf{Wrong: } If the ground truth label is the opposite of the predicted label (i.e. predicting an image from a tumor tissue as normal, or predicting an image from a normal tissue as tumor)
    \item \textbf{Non-Diagnostic: } If the predicted label is non-diagnostic
\end{itemize}

As shown in Table~\ref{fig:classifier}, the restored images have a significantly higher ``correct" rate while having a much lower percentage of misclassified or non-diagnostic images. This means that RSCD can restore a significant portion of the images that were originally classified as non-diagnostic or wrongly classified due to poor quality, and make them high-quality enough to be correctly classified.

\begin{figure*}
    \centering
    \includegraphics[width=\textwidth]{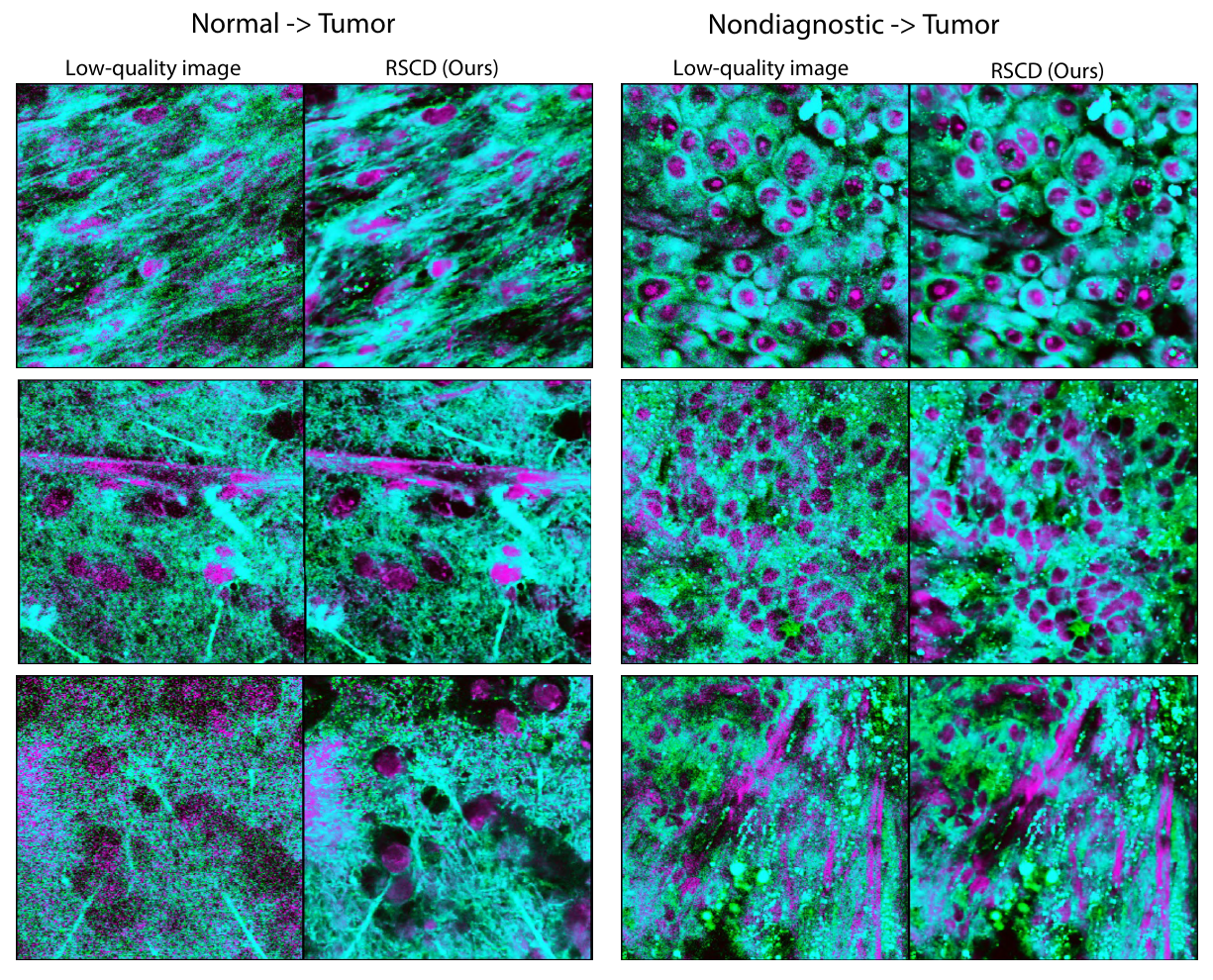}
    \caption{Examples of RSCD improving automated diagnosis. \textbf{Left}: Panel of images before and after restoration. The classification model \cite{Hollon2020-ez} incorrectly classified low-quality images of tumor tissue as normal brain tissue prior to image restoration. Following image restoration with RSCD, images were correctly classified as tumor tissue. Incorrect normal classification errors most commonly arise when tumor cells infiltrate a normal brain, which is the most common growth pattern for diffuse gliomas (shown here). \textbf{Right:} Panel of images classified as non-diagnostic prior to image restoration. Interestingly, the classification model can be sensitive to \emph{even small amounts of image degradation}. In this clinical context, image degradation functions as adversarial noise with respect to brain tumor classification. Following image restoration with RSCD, the model correctly classifies the image as tumor tissue.}
    \label{fig:diagnosisexample}
\end{figure*}

We include examples of original/restored images with their automated diagnosis results in Figure~\ref{fig:diagnosisexample}.

\subsection{Z-stack Image Restoration}
\label{app:zstack}

\paragraph{Image acquisition.}
Non-destructive z-stack SRH imaging often results in noisy images in deeper slices due to tissue scattering from the structures in the slices above, especially in specimens that contain blood or necrotic tissue. Z-stack SRH imaging follows the protocol for normal SRH imaging. Once the scanning of a whole slide is complete, the SRH imaging system automatically adjusts the laser focus, and an image at a higher depth is subsequently acquired. The z-stacked images are acquired starting at the default depth of 22 $\mu m$, with a z-resolution of 2 $\mu m$ for 20 z-sections (reaching a final depth of 60 $\mu m$). Each z-stacked whole-slide is then patched into $256\times 256 \times 20$ volumes for denoising.

\begin{figure*}
    \centering
    \includegraphics[width=.9\textwidth]{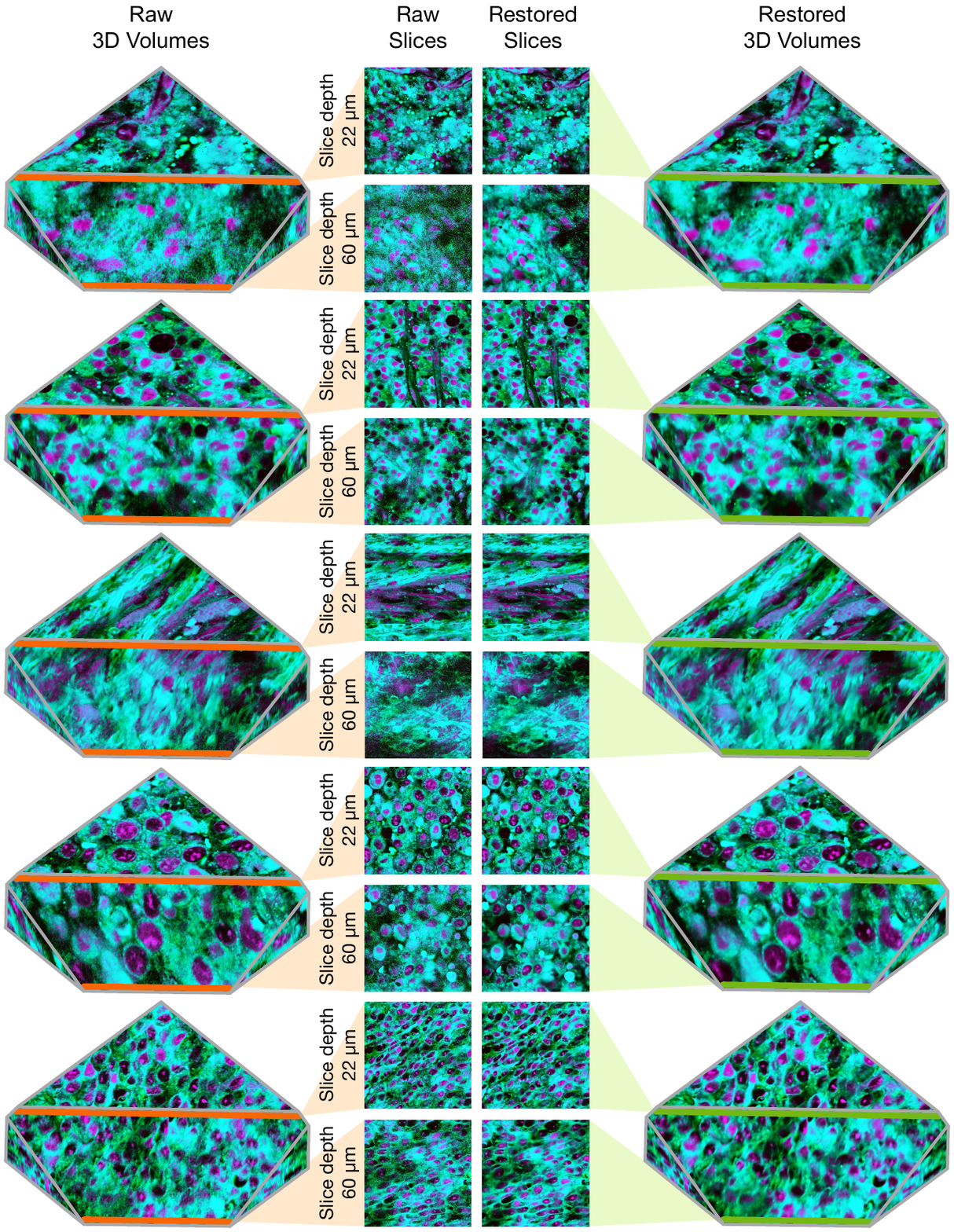}
    \caption{Additional depth z-stack restoration results. The raw 3D z-stack volumes are shown on the left and the restored volumes using RSCD are shown on the right. The center images are the cross-section views of the first and the last slices, at 22 and 60 microns, respectively. Image noise increases as the tissue depth increases due to tissue scattering. With RSCD, the restored images show little to no additional noise in deep images.}
    \label{fig:zstackadditional}
\end{figure*}

\paragraph{Additional results.} We include additional examples of restored z-stack images spanning a variety of tumor types in Figure~\ref{fig:zstackadditional}.

\end{document}